# The eddy current distortion in the multiband diffusion images: diagnosis and correction

Jiancheng Zhuang

*Dornsife Imaging Center*

*University of Southern California, Los Angeles, CA, United States*

**Abstract**

The diffusion weighted images acquired with the multiband sequence or the Lifespan protocols shows a type of slice distortion artifact. We find that this artifact is caused by the eddy currents, which can be induced by the diffusion gradient associated with either current DW image or the previous DW images. The artifact can be corrected by further tuning the compensation circuit in the MR hardware, or by a correction algorithm which includes the diffusion gradients from the current and previous DW images.

**Introduction**

Diffusion tensor imaging (DTI) has become a popular tool to provide information about the intrinsic architectures of white matter in the human brain (1). However, DTI technology has a significant limitation in resolving orientation heterogeneity within single voxels due to the constraints of tensor models. As an obstacle for efforts to construct white matter pathways from diffusion MRI data, the limitation has prompted the development of diffusion imaging methods capable of resolving intravoxel fiber crossings, such as HARDI

(high angular resolution diffusion imaging) (2). Among HARDI techniques, diffusion spectrum imaging (DSI) employs the Fourier relation between the diffusion signal and the function of diffusion wave vector q (3), and q-ball imaging (QBI) uses Funk–Radon transform to process the HARDI signal (4).

However, because most of the HARDI techniques require high to ultra-high diffusion sensitizing gradients (b>=2000 s/mm$^2$), the capability of HARDI to provide valid and reliable information about tissue structures can be affected adversely by eddy current artifacts. In echo planar images, usually used to acquire diffusion weighted (DW) images, eddy currents produce significant distortions in the phase-encoding direction because of the relatively low bandwidth in that direction and the large changes in diffusion gradients during HARDI scanning. Image distortions from eddy currents blur the interface of gray and white matter tissues, cause misregistration between individual diffusion-weighted images, produce erroneous calculations of the diffusion signals, and spoil the detected high angle-resolution characteristics of diffusion at each voxel.

Eddy current distortions can be reduced effectively in one of three ways: first, by selection of an appropriate pulse sequence (such as a dual spin-echo sequence) (5, 6) or gradient waveforms (such as bipolar gradients) (7); second, by correction of k-space data, such as calibration of eddy current artifacts in k-space (8, 9, 10); third, by post-acquisition image processing that registers diffusion-weighted images to reference images. This third approach, based on post-processing algorithms, is appealing because of its relative ease and accessibility. One widely used post-processing algorithm, Iterative Cross-Correlation (ICC) (11), estimates distortions in DW images by cross-correlating them with an undistorted baseline image in terms of scaling, shear, and translation along the phase-encoding direction. The estimated distortion parameters are then used to correct all distorted images (11, 12, 13, 14).

One serious limitation of the original ICC algorithm (11), however, is its inability to correct image distortions at high b-values. The contrasts of cerebrospinal fluid (CSF), gray matter, and white matter in images acquired with no diffusion weighting, differ greatly from the contrasts found in images acquired with high (b-value) diffusion weighting. The contrast differences lead to unreliable registration of the two types of images, which in turn interferes with eddy current distortion corrections. This problem is more acutely felt in most q-space diffusion images for which high or ultra-high b values are commonly used (2, 3, 4).

Various methods have been proposed to more accurately estimate the distortion effects of eddy currents. Some investigators proposed a method of extrapolating distortion parameters from low to high b-value images (11). Others employed the ICC algorithm with reference to CSF-suppressed images (such as FLAIR) to minimize the major source of contrast change in images acquired with different b-values (15). Diffusion weighted images of a water phantom have also been used to measure distortion parameters directly, and these parameters can then be used to calibrate the ICC of brain images (13). Although these procedures extend the possibility to use the ICC algorithm with b-values as high as 1000 s·mm$^{-2}$, they require acquisition of additional images that prolongs scanning times, which is not always desirable.

Two recent approaches use only diffusion-weighted images to estimate relative distortions. One approach, co-registration of pairs of DW images with exactly the reversed diffusion gradients followed by corrections of the distortions using ICC, will double the acquisition time (14); another, applying the known gradient strength and direction to model the absolute distortions only between DW images, may involve inaccurate image co-registrations, especially at ultra-high diffusion gradient strength, due to image contrast differences resulting from the change of diffusion gradient directions (16, 17).

The recently developed multilband slice-accelerated technique can provide high angular resolution for measuring the diffusion signals in shorter time length than conventional EPI sequence (18), especially in Human Connectome Lifespan projects. However, how the eddy current specially affects the multiband diffusion images or whether the multiband acquisition is special on the characters of eddy current is still under investigation.

Here we describe a new method to detect eddy current distortions and model the distortion with the known x, y, and z components of diffusion gradients exclusively from DW images with current and previous diffusion gradient directions. The algorithm was validated in the experimental phantom data. Finally, we demonstrate its successful application to correct distortions in diffusion weighted images of the human brain.

# Methods and Results

## A. Inspection

One fBIRN phantom and two adults with no history of neurological disease were scanned on a Siemens 3T Trio Prisma MRI system (Siemens Healthcare, Erlangen, Germany) at our institute. HARDI data were acquired using multiband DWI sequence with the following parameters as in the Lifespan projects: repetition time (TR) 3222 msec, echo time (TE) 89.2 msec, 197 diffusion gradient directions, 92 axial slices for whole brain coverage, MB factor as 4, and maximum b value as 2000 s/mm². As in the result images (Figure 1), an artifact shows up as misalignment between slices in both the diffusion weighted images and b0 images in both human and phantom data.

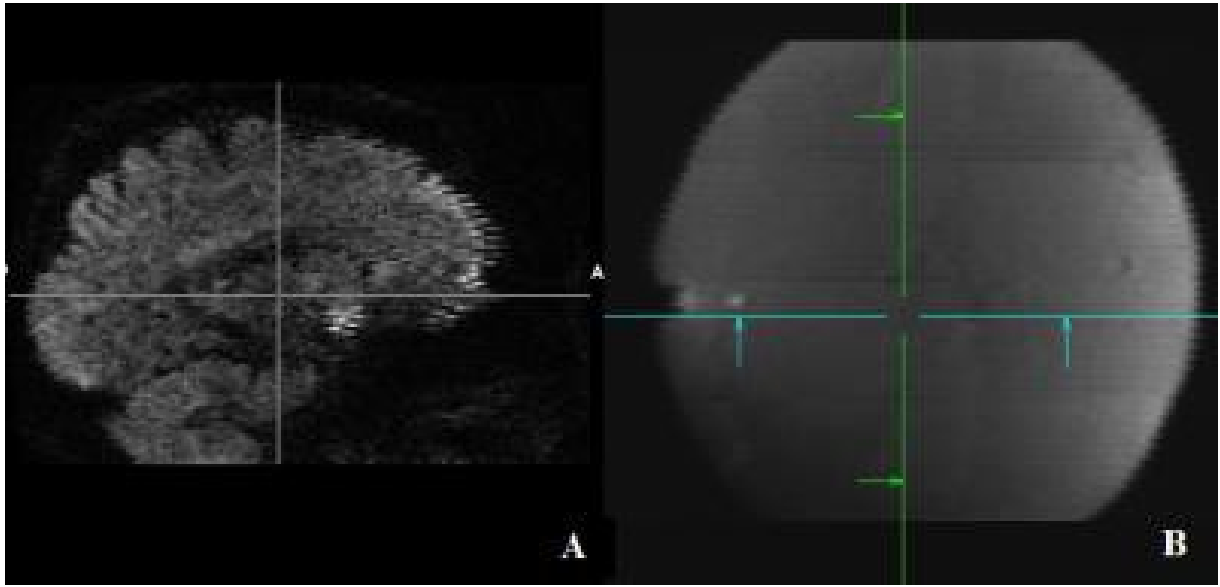

Figure 1. The slice distortion artifact shown on a human brain (A) and a phantom (B).

## B. Diagnosis

To diagnose the possible reason of this artifact, we change the slice order from interleaved into ascending, and switch the phasing encoding direction into AP, PA and RL in the DWI scans on the fBIRN phantom. The results show that the PA and AP phase encodings reverse the slice stretch and shrink at the sagittal view, and RL phase encoding makes the slice shift at the coronal view (Figure 2). It is a typical eddy current effect. However, this kind of eddy current artifact is rarely seen on the b0 images in the diffusion scans and in the conventional EPI images. The further diagnosis on the eddy currents on the MRI is done on the next.

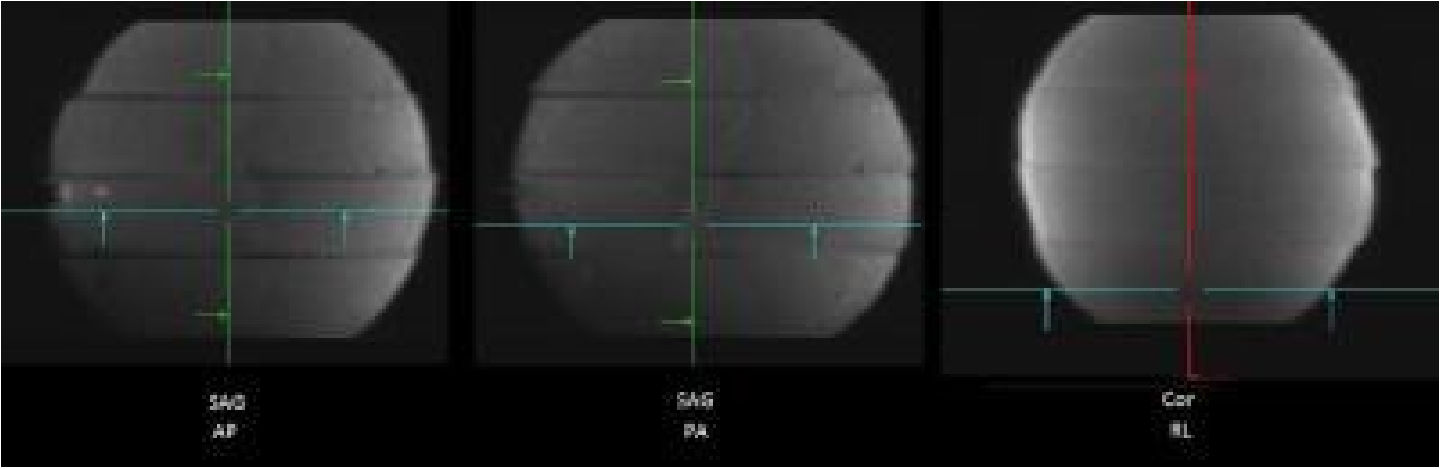

Figure 2. The phantom data acquired with ascending slice order, and AP, PA and RL phase encoding.

*C. Hardware correction*

Using Siemens service hardware, we map the eddy current along the time axis (Figure 3A). We found the typical compensation of eddy current mainly focused on the time range of 2-100ms, but the amplitude of eddy current around 1-4 seconds is still large. This time range is close to the TR (3222 ms) used in the Lifespan protocol. The super short TR and super high diffusion gradient strength in this MB protocol make it possible that the eddy current affects different images across TR. That is why the b0 image in the Lifespan protocol displays this artifact.

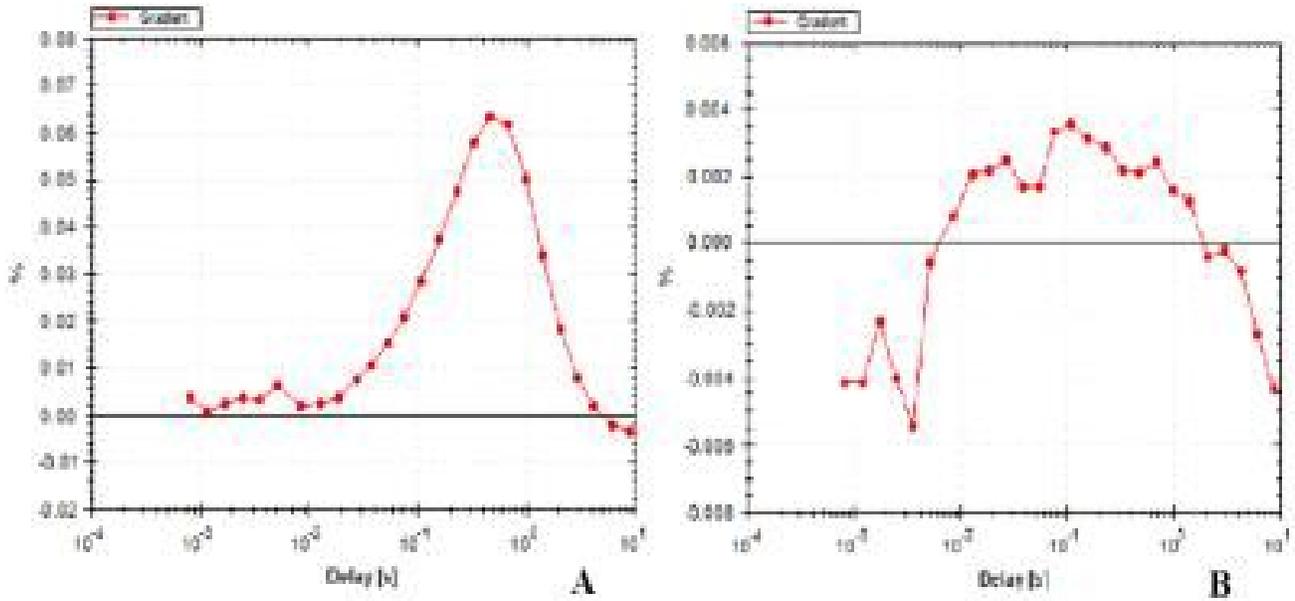

Figure 3. Eddy current vs time before (A) and after (B) further tuning the compensation circuits.

Before multiband sequence is available, most diffusion scans of the whole brain have TR much longer by a few seconds, so this eddy current effect from previous volume is very low in the conventional EPI images. After our re-tuning of the compensation circuits, the eddy current after the specified time range (1- 4s) is also reduced (Figure 3B). Therefore this artifact is deducted significantly in the new images (Figure 4B).

*D. Simulation of the distortion*

Because diffusion-sensitizing gradients consist of components along each of the x, y, and z axes, these gradient components should induce eddy currents independently. The eddy currents induced by a change in a single gradient component, the x gradient for an example, can be distributed along the x, y, and z axes. Such eddy currents produce residual gradient fields in the frequency encoding, phase encoding, and slice-selection directions. These residual gradients in turn cause shearing, scaling, and translational distortions, all visible along the phase encoding direction of the EPI images (8). Assuming that the interaction between these three components of gradient fields is negligible (11), then the total distortion from eddy currents will equal the linear sum of the three components of the distortion induced by the x, y and z gradients.

Accordingly, the x, y and z components of the *i*-th diffusion gradient $\mathbf{G}_i = (G_{ix}, G_{iy}, G_{iz})$ will produce the corresponding image *translation* distortion $\mathbf{G}_i \cdot \mathbf{T} = G_{ix}T_x + G_{iy}T_y + G_{iz}T_z$, where $\mathbf{T} = (T_x, T_y, T_z)$ are the translations along the phase encoding direction induced by the corresponding unit changes in the x, y, and z gradients. As shown in above, the distortion at *i*-th DW image is also affected by the previous *i-1*-th diffusion gradient as $\alpha \mathbf{G}_{i-1} \cdot \mathbf{T}$, where α is the decay factor of eddy current from $\mathbf{G}_{i-1}$ at the time point of *i*-th DW image. The resulting distortion in translation $D_{ti}$ from the alignment between the images of the *i*-th diffusion gradient direction and the *1st* gradient direction can be calculated for $i > 1$ as,

$$D_{ti} = \mathbf{G}_i \cdot \mathbf{T} + \alpha \mathbf{G}_{i-1} \cdot \mathbf{T} - \mathbf{G}_1 \cdot \mathbf{T},$$

or

$$\mathbf{G}' \cdot \mathbf{T} = \mathbf{D}_t, \quad [1]$$

where the rows of matrix $\mathbf{G}'$ are formed by $(G_{ix}+\alpha G_{i-1\,x}-G_{1x},\ G_{iy}+\alpha G_{i-1\,y}-G_{1y},\ G_{iz}+\alpha G_{i-1\,z}-G_{1z})$ for the *i*-th diffusion gradient, and $\mathbf{D}_t$ is the distortion vector of image translations that is measured by the registration between the image from the *i*-th diffusion gradient and the image from the first diffusion gradient. The three unknown elements of vector $\mathbf{T}$ can be calculated as

$$\mathbf{T} = (\mathbf{G}'^T \cdot \mathbf{G}')^{-1} \cdot \mathbf{G}'^T \cdot \mathbf{D}_t, \quad [2]$$

where the superscripts "*T*" and "–1" denote matrix transposition and inversion, respectively.

Similarly, a vector $\mathbf{S}$ of the *shear* distortion induced by a unit change of the x, y, and z components of the gradient can be calculated using equations

$$\mathbf{S} = (\mathbf{G}'^T \cdot \mathbf{G}')^{-1} \cdot \mathbf{G}'^T \cdot \mathbf{D}_s, \quad [3]$$

where $D_s$ is the vector for shearing, which is measured in the registration of the image from the *i*-th diffusion gradient with the image from the first diffusion gradient.

Finally, the *scaling* (or *magnification*) distortion $D_{mi}$, measured by comparing the image from the *i*-th diffusion gradient and the image from the first diffusion gradient, is calculated for $i > 1$ as

$$D_{mi} = \frac{1 + G_{ix}M_x + G_{iy}M_y + G_{iz}M_z + \alpha(G_{ix}M_x + G_{iy}M_y + G_{iz}M_z)}{1 + G_{1x}M_x + G_{1y}M_y + G_{1z}M_z},$$

where $M_x$, $M_y$, and $M_z$ are the unknown components of scaling induced by unit changes in the x, y, and z of the gradient components, respectively. The following can therefore be derived:

$$\mathbf{G}'' \cdot \mathbf{M} = \mathbf{D}'_m, \qquad [4]$$

where the matrix $\mathbf{G}''$ is formed as $(G_{ix}+\alpha G_{i-1\,x}-D_{mi}G_{1x},\ G_{iy}+\alpha G_{i-1\,y}-D_{mi}G_{1y},\ G_{iz}+\alpha G_{i-1\,z}-D_{mi}G_{1z})$ and $D'_{mi} = D_{mi} - 1$ for $i>1$. The vector $\mathbf{M} = (M_x, M_y, M_z)$ can thus be obtained as

$$\mathbf{M} = (\mathbf{G}''^T \cdot \mathbf{G}'')^{-1} \cdot \mathbf{G}''^T \cdot \mathbf{D}'_m. \qquad [5]$$

Given the model parameters for the distortions $\mathbf{T}$, $\mathbf{S}$, and $\mathbf{M}$, we can determine the total distortions for the $i$-th diffusion gradient in relation to the undistorted, non-DW images using the dot products of $\mathbf{G_i} \cdot \mathbf{T}$, $\mathbf{G_i} \cdot \mathbf{S}$, and $\mathbf{G_i} \cdot \mathbf{M}$. Thereafter image distortions can be corrected by reverse application of these parameters to the distorted DW images.

### E. Software correction

To correct the diffusion weighted images which already has this artifact, we apply the algorithm for correcting the eddy current distortions using the known diffusion gradient strengths and directions, as in the above equations [2, 3, 5]. Because the distortion in one image is not only affected by the current diffusion gradient but also by the previous diffusion gradient, the algorithm should include the diffusion gradient associated with the previous diffusion weighted volume. After correction by this way, the artifact is reduced to the minimal level (Figure 4A).

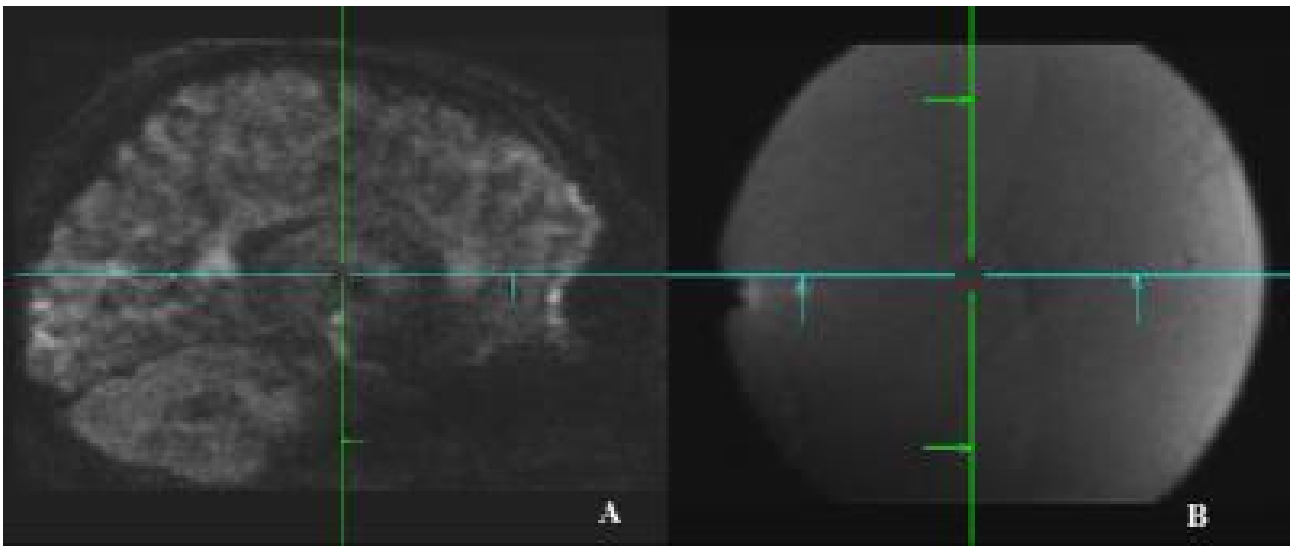

Figure 4. The corrected images on human brain (A) and phantom (B).

## Conclusion

The slice distortion shown in the diffusion weighted images acquired with the multiband sequence or the Lifespan protocols is caused by the eddy currents, which can be induced by the diffusion gradient associated with either current DW image or the previous DW image. It can be corrected by further tuning the hardware compensation, or by a correction algorithm which includes the diffusion gradients from the previous DW image.